\documentclass[conference]{IEEEtran}

\usepackage{paper}

\begin{document}
\title{Formal Verification for Deep Learning-based Power Control in Massive MIMO\\
    \author{
    \IEEEauthorblockN{
    Thanh Le\IEEEauthorrefmark{1},
    Takeshi Matsumura\IEEEauthorrefmark{1},
    Yusheng Ji\IEEEauthorrefmark{2},
    John C.S. Lui\IEEEauthorrefmark{3}
    }
    \IEEEauthorblockA{\IEEEauthorrefmark{1}National Institute of Information and Communications Technology (NICT), Japan}
    \IEEEauthorblockA{\IEEEauthorrefmark{2}National Institute of Informatics (NII), Japan, \IEEEauthorrefmark{3}The Chinese University of Hong Kong (CUHK), China}
    }

}
\maketitle

\begin{abstract}
    Deep learning is a promising approach to optimize wireless communication by simplifying the search for near-optimal solutions.
    Prior studies on deep learning–based wireless communication optimization have explored supervised learning approaches that map raw user information,
    such as location or channel state information, to optimal power allocation vectors.
    While this approach demonstrates competitive performance, it is susceptible to adversarial attacks via input perturbations.
    Current defense mechanisms primarily rely on empirical methods, which do not provide formal guarantees of robustness.
    We fill this gap by proposing a formal verification framework to evaluate the robustness of deep learning-based power allocation in
    multi-cell massive multiple-input multiple-output (MIMO) systems against a wide range of potential adversarial input manipulations.
    To the best of our knowledge, this is the first attempt to formally verify deep neural networks in a regression setting with non-linear output constraints.
    We model the adversary’s capabilities using hyper-rectangle constraints on their perturbation,
    adopt the abstraction-based bound-propagation technique (DeepPoly) to bound the interval of potential allocated powers,
    and formulate the minimum performance requirements as a constrained program for numerical feasibility analysis.
    Evaluation on publicly available datasets for power allocation in multi-cell massive MIMO indicates that
    a well-trained model can guarantee the local robustness under location perturbation by $\pm$ 1m while retaining a maximum 1\% optimality gap. 
\end{abstract}


\glsresetall
\begin{NoHyper}
\section{Introduction}
    \label{sec:introduction}
    Massive \gls{mimo} technology is a key component of next-generation wireless networks,
    designed to accommodate the growing number of connected \gls{ue} and their diverse demands for new services and applications \cite{bjornson2017massive}.
    By deploying a large number of antennas at the \gls{bs}, massive \gls{mimo} can leverage multiplexing gains and enable high data rates.
    In multi-cell massive \gls{mimo} networks, optimizing power control is crucial
    for achieving high performance and reducing interference between \gls{ue}s both within a cell and across different cells.

    Power control in massive \gls{mimo} increasingly leverages \gls{dnn} to reduce complexity by inferring near-optimal solutions within $O(1)$ time complexity
    while requiring less engineering effort to design new heuristics if the problem formulation changes.
    Exciting developments have emerged in this area,
    \eg, \gls{dnn} has been supervised to approximate optimal power allocation algorithms in massive \gls{mimo} systems \cite{sanguinetti2018deep}.
    This work utilizes the ability of \gls{dnn} to approximate arbitrary functions to map the positions of \gls{ue}s in the multi-cell network to their optimal power allocation.
    A few following studies have focused on training distributed architectures using only local information within each cell \cite{chakraborty2019centralized,zaher2023learning}.
    Another work has addressed varying numbers of users by utilizing \gls{cnn} \cite{van2020power},
    while deep reinforcement learning has been applied to control power in massive \gls{mimo} networks with user mobility \cite{zhang2025multi}.

    However, \gls{dnn}-based power allocation is susceptible to adversarial attacks,
    where small perturbations can be added to inputs and manipulated \gls{dnn}s into allocating non-optimal power to \gls{ue}s \cite{manoj2021adversarial,santos2021universal}.
    These perturbations are small enough to evade anomaly detection system but carefully crafted to significantly impair the \gls{dnn}'s performance,
    \eg, reducing transmission rates, fairness, and degrade quality of service.
    Previous work \cite{manoj2021adversarial} proposed an attack on \gls{dnn}-based power control with the input being the locations of \gls{ue}s.
    By using \gls{gnss} spoofing techniques,
    attackers can alter the positions of nearby \gls{ue}s,
    thereby causing infeasible and non-optimal power allocations.

    To the best of our knowledge, there are three main defense approaches:
    (1) denoising autoencoders, (2) adversarial training, and (3) \gls{nnv}.
    Firstly, denoising autoencoder networks can be employed to rectify malicious inputs~\cite{sahay2023defending};
    however, this incurs additional computational overhead during inference.
    Secondly, adversarial training mixes sampled adversarial inputs to the training dataset for retraining~\cite{manoj2022downlink}.
    However, given the real-valued, high-dimensional input space, the sampled adversarial inputs do not guarantee completeness.
    Besides, this type of non-certifying defense was found to be evaded by stronger adversarial attack schemes~\cite{athalye2018obfuscated}.
    To mitigate this cat-and-mouse game, there is a gaining traction in defenses with \gls{nnv}~\cite{singh2019abstract,cohen2019certified,xu2020automatic},
    as \gls{nnv} methods provide absolute robustness guarantee.

    General \gls{nnv} techniques have primarily focused on verifying the safety of computer vision~\cite{duong2026verifying,duong2026verifying2}, natural language processing, and goal-oriented semantic communication~\cite{le2026verifying}, most of them are classification or unconstrained regression tasks~\cite{le2026formal}.
    In contrast, our problem involves verifying non-linear performance guarantees,
    \eg, guarantee minimum product of \gls{sinr},
    which cannot be directly addressed by general \gls{nnv} techniques and require additional post-processing.
    In \gls{dnn}-based for wireless communication domain,
    a previous work verified robustness of \gls{dnn}-based antenna selection for massive \gls{mimo} systems~\cite{kim2025certified},
    in which \gls{dnn}s are trained to classify channel information to optimal antenna configurations.
    As far as we know, verifying \gls{dnn} in regression-based tasks with non-linear throughput and fairness constraints remains an open challenge.

    We fill this gap by proposing a verification framework for \gls{dnn}-based power control in massive \gls{mimo} systems.
    Our proposed framework systematically analyzes whether the trained \gls{dnn} meets performance requirements under various levels of adversarial perturbations.
    First, we convert different levels of adversarial perturbation into a wide variety of hyper-rectangle input properties.
    We apply abstraction-based bound-propagation \gls{nnv} technique called DeepPoly \cite{xu2020automatic,singh2019abstract} to compute an over-approximation of the output,
    which covers all possible power allocations.
    Then, we propose an adaptation of the existing \gls{nnv} to work with the optimization goal in the considered problem.
    Particularly, we check whether the model satisfies a guarantee on product of \gls{sinr},
    \eg, the optimality gap must fall below a certain threshold,
    by formulating a constrained program for feasibility analysis.


\section{Preliminaries}
    \label{sec:preliminary}
    \subsection{System Model}
        To investigate the robustness of deep learning-based power control in a multi-cell massive \gls{mimo} networks,
        we reproduce the system model and \gls{dnn}-based power control method which have been made publicly available in \cite{sanguinetti2018deep}.
        In this system, the massive \gls{mimo} network consist of $J$ cells, each cell having a \gls{bs} of $M$ antennas and covers $K$ \gls{ue}s.
        The downlink signal transmitted by the \gls{bs} in cell $j$ is given by
        $v_j = \sum_{k=1}^K \mathbf{w}_{jk} u_{jk}$,
        where $u_{jk} \in \mathcal{N}(0, \rho_{jk})$ is the data signal in the downlink for \gls{ue} $k$ in \gls{bs} $l$,
        with a precoding vector $w_{jk} \in \mathcal{C}^M$ that determines the transmission beamforming satisfying $||w_{jk}||^2=1$,
        thus, $\rho_{jk}$ is the transmission power.
        The best precoding schemes that have been adopted by \cite{sanguinetti2018deep} in generating dataset is \gls{mmmse}.
        The downlink \gls{sinr} for \gls{ue} $k$ in cell $j$ is:
        \begin{equation}
            \gamma_{jk} = \frac{\rho_{jk} a_{jk}}{\sum_{l=1}^J \sum_{i=1}^K \rho_{li} b_{lijk} + \sigma^2},
        \end{equation}
        where $\sigma^2$ is the \gls{awgn} variance and the average channel gain is
        $a_{jk} = |\mathbb{E}[{\mathbf{w}_{jk}^H \mathbf{h}_{jk}^j}]|^2$ ,
        with $\mathbf{h}_{jk}^j$ denoting the channel between \gls{bs} $j$ and \gls{ue} $k$ in cell $j$.
        The average interference gain is:
        \begin{equation}
            b_{lijk} =
            \begin{cases}
                \mathbb{E}[|\mathbf{w}_{li}^H \mathbf{h}^{l}_{jk}|^2] - |\mathbb{E}[{\mathbf{w}_{li}^H \mathbf{h}^{l}_{jk}}]|^2, \text{ if } (l, i) = (j, k), \\
                \mathbb{E}[|\mathbf{w}_{jk}^H \mathbf{h}^{l}_{jk}|^2], \text{ otherwise.}
            \end{cases}
        \end{equation}

\begin{figure*}[h]
    \centering
    \includegraphics[width=0.9\linewidth]{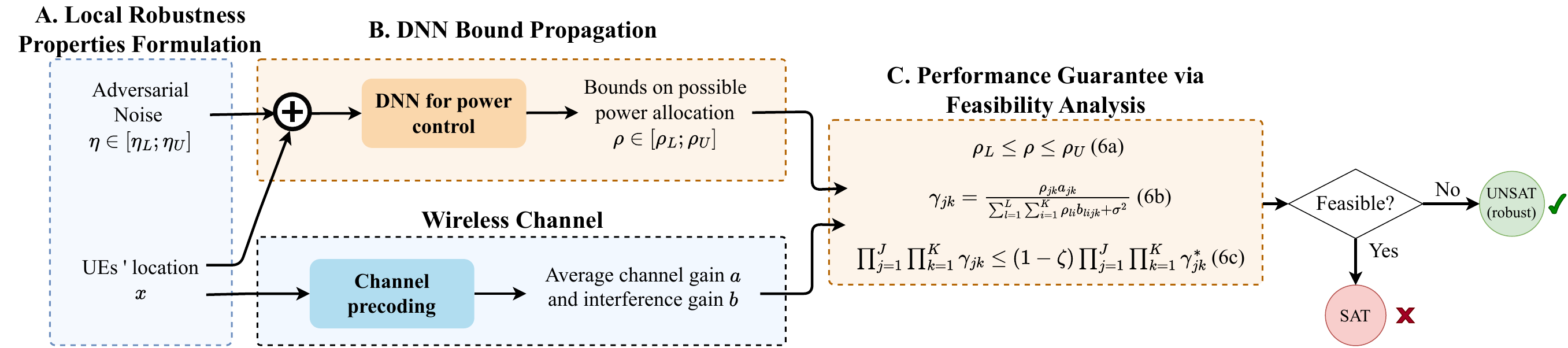}
    \caption{Overview of the proposed \gls{nnv} framework to verify \gls{dnn}-based power control in multi-cell massive \gls{mimo} networks.}
    \label{fig:framework}
\end{figure*}

    \subsection{Power Control Problem}
        In this study, we attempt to verify the \gls{dnn}-based optimal power allocation strategy to maximize the max-product \gls{sinr} \cite{sanguinetti2018deep}, which is formulated as:
        \begin{subequations} \label{eq:power-allocation-problem}
            \begin{align}
                \underset{\rho_{jk}: \forall j, k}{\text{maximize}} \quad & \prod_{j=1}^{J} \prod_{k=1}^K \gamma_{jk}, \label{eq:power-allocation-problem-a} \\
                \text{subject to} \quad & \sum_{k=1}^K \rho_{jk} \le P_{\text{max}}, \quad j \in [1, J]. \label{eq:power-allocation-problem-b}
            \end{align}
        \end{subequations}
        The problem above seeks to maximize objective in \autoref{eq:power-allocation-problem-a}, which is the product of \gls{ue}s' \gls{sinr}, while \autoref{eq:power-allocation-problem-b} ensures that the allocated power adhere to the maximum power threshold $P_{max}$ promoting a trade-off between \gls{ue}'s performance and total throughput.
        Compared to uniform max power control and max-min spectral efficiency power control, max product \gls{sinr} offers a balance between fairness and efficiency~\cite{sanguinetti2018deep}.

    \subsection{Deep Learning-based Power Control}
        In \cite{sanguinetti2018deep}, \gls{dnn}s are employed to compute an approximated optimal power allocation that maximizes the \gls{sinr} product of all \gls{ue}s.
        A publicly available dataset\footnote{\url{https://data.ieeemlc.org/Ds2Detail}} is provided as $\{x^{(n)}, \rho^{(n)}\}_{n=1}^{N_T}$, where $x^{(n)} \in \mathbb{R}^{J \times K \times 2}$ represents an input feature contains location of \gls{ue}s, the output label $\rho^{(n)} \in [0, \rho_{max}]^{J \times K}$ is the optimal power, and $N_T$ denotes the size of the dataset.
        We denote a regression-based \gls{dnn} model as $N(.; \theta)$, with $x$ as the input to the model and the predicted output as $N(x; \theta) = \rho$, where $\theta$ constitutes the set of parameters of the model $N$.
        The corresponding loss function is represented as $\mathcal{L}(x, \rho; \theta)$.
        The goal of the \gls{dnn} model is to learn the mapping from the \gls{ue}s' geographical positions, $x$, to the optimal power allocation solution, $\rho$ produced by mathematical optimization tools.
        A feed-forward neural network architecture and \gls{mse} loss function that has been adopted to approximate the optimal power allocation.
        In the standard setting outlined in \cite{sanguinetti2018deep}, the input $x$ is a vector representing the geographical locations of all \gls{ue}s in the system, while $\rho$ is a vector containing the optimal power solution obtained by solving the power allocation problem (refer to \autoref{eq:power-allocation-problem})


    \subsection{\gls{nnv} - An Overview}
        Given a \gls{dnn} $N$ and a property $\phi$, the general \emph{\gls{nnv} problem} asks if $\phi$ is a valid property of $N$.
        Typically, $\phi$ is a formula of the form $\phi_{in} \implies \phi_{out}$, where $\phi_{in}$ is a property over the inputs of $N$ and $\phi_{out}$ is a property over the outputs of $N$.
        This property has been used to encode safety and security requirements of \gls{dnn}~\cite{duong2025neuralsat}.
        A \gls{dnn} verifier attempts to find a \emph{counterexample} to $N$ that satisfies $\phi_{in}$ but violates $\phi_{out}$.
        If no such counterexample exists, $\phi$ is an unsatisfiable property of $N$ (\texttt{UNSAT} or proven robust); otherwise, satisfiable (\texttt{SAT}).

\section{Proposed Verification Framework}
\label{sec:propose}

    \autoref{fig:framework} illustrates our proposed robustness guarantees framework for \gls{dnn}-based power control in massive \gls{mimo} through a three-phase pipeline.
    First, we formulate local robustness input properties encapsulating possible adversarial noise around system inputs.
    Then, an abstraction-based bound-propagation \gls{dnn} verifier method \cite{xu2020automatic,singh2019abstract} is adopted to compute bounds on possible power allocations provided by a trained \gls{dnn}.
    Finally, given the power allocation bounds, we formulate a constrained program to prove the non-existence of a power allocation that violates the maximum optimality gap guarantee, \eg, reducing the optimization objective by more than $\zeta$.
    \al{The three modules/phases should match with the subsection's names below (currently the phase name in Fig.2 are "Input Property,...".}
    \tl{--- done ---}
    \al{Also, the purpose of each phase should be shortly mentioned here (or at the beginning of each subseccion). }

    \subsection{Local Robustness Properties Formulation}
        We first define the input properties encoding the local robustness requirements of \gls{dnn}-based power allocation.
        Following the threat model in \cite{manoj2021adversarial,manoj2022downlink}, we assume that an attacker can introduce controlled perturbations to the input fed into the \gls{dnn}.
        For instance, the attacker may manipulate the reported positions of \gls{ue}s using spoofing techniques, in which global navigation satellite system receivers are deployed near the users’ actual locations.
        \al{I think that at this, you can directly refer to the input of DNN-based MIMO power control problem (i.e., $x^{(n)}$?).}
        \tl{--- sure, I made several changes in the following sentences by simply mention input ---}
        Furthermore, the perturbation remains minimal relative to the actual input $x^{(n)}$.
        The adversary’s objective is to compute input perturbations along the gradient direction to maximize the loss function (e.g, maximize \gls{mse} loss to optimal power allocation), thereby degrading the performance of the \gls{dnn}-based power allocation system.

        To formally verify the robustness of the considered \gls{dnn}-based power allocation scheme, we define input specifications $\phi_{in} := [\eta_L, \eta_U]$, such that $\eta_L$ and $\eta_U$ are two vectors having the same size as input $x^{(n)} $, to encode permitted perturbation levels.
        \al{$\eta_L, \eta_U$ should be defined more clearly. What do you mean perturbation levels? are $\eta_L, \eta_U$ discrete?}
        \tl{--- I clarified $\eta_L, \eta_U$ space---}
        The adversary seeks to generate an adversarial example $x'$ within a constrained neighborhood of $x$ such that the \gls{dnn} output becomes erroneous.
        This adversarial example is constructed by adding a bounded perturbation $x' = x^{(n)} + \eta$ to the original input $x^{(n)}$, where $\eta \in [\eta_L, \eta_U]$. 
        Similarly, output specifications $\phi_{out}$ are introduced to capture minimum performance guarantees, requiring that the achieved performance remains at least $1 -\zeta$ times that of the optimal solution under controlled conditions.
        This property, referred to as local robustness, ensures stability against small adversarial perturbations around given inputs.

    \subsection{DNN Bound Propagation}
        In this phase, we aim to bound the upper and lower limits of power allocations given the input property and the trained \gls{dnn}.
        Bounding the output range of a \gls{dnn} under constrained input perturbations has emerged as a critical topic in \gls{nnv} and certified adversarial defense~\cite{xu2020automatic,singh2019abstract}.
        For a neural network $N(x, \theta)$, we analyze its behavior at a nominal input $x^{(n)}$ subject to a bounded perturbation $\eta$, where $x = x^{(n)} + \eta$ and $x$ lies within an $l_p$-norm ball $\mathcal{B}_p(x^{(n)}, r)$.
        Computing exact output bounds of a \gls{dnn} is generally intractable~\cite{duong2026compositional,duong2025neuralsat2}.

        To address this challenge, we employ DeepPoly~\cite{singh2019abstract} abstraction,
        which provides provable linear relaxation of bounds on neuron outputs given specified input perturbations.
        This method constructs two linear functions that serve as guaranteed lower and upper bounds for each neuron’s output with respect to its perturbed input. 
        The procedure begins by relaxing non-linear activation functions into linear constraints.
        These relaxations of bounds on output of non-linear activations are then propagated through the network architecture, layer by layer, from the input to the final output.
        DeepPoly accepts input specifications consisting of bounds $[x^{(n)} + \eta_L, x^{(n)} + \eta_U]$, along with the trained \gls{dnn} in \gls{vnncomp} format~\cite{kaulen20256th}.
        The output is a set of tight bounds $[\rho_L, \rho_U]$ on the \gls{dnn}s' outputs.

        Note that
        \gls{nnv} has been generalized to general computational graphs \cite{xu2020automatic}, enabling the computation of linear output bounds not restricted to feed-forward neural networks but also able to find bounds of modern architectures, \eg, \gls{cnn}s, ResNets, or Transformers.
        However, we will adhere to the feed-forward neural network architecture used in publicly available implementations to ensure reproducibility and consistency with previous works~\cite{sanguinetti2018deep,manoj2021adversarial}.

    \subsection{Performance Guarantee via Feasibility Analysis}
        The performance guarantee of \gls{dnn}-based power control in multi-cell massive \gls{mimo} networks cannot be captured by existing \gls{nnv} properties, which primarily target classification tasks or generic regression objectives with linear constraints~\cite{kaulen20256th}.
        We propose a constrained program to encode the non-linear performance guarantee as follow:
        \begin{subequations} \label{eq:feasibility-analysis}
            \begin{align}
                \rho_L &\le \rho \le \rho_U, \label{eq:feasibility-analysis-a} \\
                \gamma_{jk} &= \frac{\rho_{jk} a_{jk}} {\sum_{l=1}^J \sum_{i=1}^K \rho_{li} b_{lijk} + \sigma^2}, \label{eq:feasibility-analysis-b} \\
                \prod_{j=1}^J \prod_{k=1}^K \gamma_{jk} &\le (1 - \zeta) \prod_{j=1}^J \prod_{k=1}^K \gamma^{*}_{jk},  \label{eq:feasibility-analysis-c}
            \end{align}
        \end{subequations}
        where \autoref{eq:feasibility-analysis-a} constrains the power allocation variables $\rho$ to lie within the bounds $[\rho_L, \rho_U]$ obtained from the \gls{dnn} bound propagation, \autoref{eq:feasibility-analysis-b} defines the \gls{sinr} $\gamma_{jk}$ for each \gls{ue} $k$ in cell $j$ based on the power allocation and channel conditions, and \autoref{eq:feasibility-analysis-c} encodes the performance guarantee by ensuring that the product of all \gls{sinr} values does not fall below $(1-\zeta)$ times the optimal product $\gamma^{*}_{jk}$.
        Feasibility of this program is verified using numerical optimization tools.
        If there exists no feasible solution then the verification problem is deemed to be unsatisfiable or \texttt{UNSAT}.
        Consequently, the \gls{dnn} can guarantee to achieve a relative optimality gap of $\zeta$ for all adversarial inputs within $[\eta_L, \eta_U]$.
        Otherwise, the verification problem is satisfiable (\texttt{SAT}) and there may exist an exploitation.

        Since the output bounds $[\rho_L, \rho_U]$ represent an over-approximation, and the feasibility analysis in \autoref{eq:feasibility-analysis} check for violation in the superset of all possible values of $\rho$, the proof of robustness is a sound proof.
        Soundness ensures that the proposed verification framework will never produce any false positive.
        If a \gls{dnn} is claimed to be robust against an input bound $[\eta_L, \eta_U]$, then the claim is guaranteed to be true.

\section{Evaluation Results}
\label{sec:experiment}

    \subsection{Evaluation Setups}
        \subsubsection{Dataset}
            We verify the robustness of the \gls{dnn}-based power control models for multi-cell massive \gls{mimo} networks\footnote{The Python implementation of our proposed framework is available at \url{https://github.com/thanhlexyz/verify_cfmimo}}, which are trained using the publicly available dataset~\cite{sanguinetti2018deep}.
            We then extract a training set of $329,000$ pairs of \gls{ue} locations and their optimal power allocations, and another $500$ samples form the test dataset, which is independent of the training dataset~\cite{manoj2021adversarial}.
            The dataset is generated with $J=4$ cells, with each cell covering a square area of $250 \times 250$ m.
            A wrap-around topology is used to better represent interference for \gls{ue}s in the edges of the network.
            Within each cell, there are $K=5$ \gls{ue}s at a randomly and uniformly distributed location, and at a distance that is larger than 35m from the \gls{bs}.
            The bandwidth is $ B=20$ MHz, with the total receiver noise power $\sigma=-94$ dBm.
            The pilot reuse factor $\tau_p=1$, and the maximum transmit power per \gls{ue} is $20$ dBm.

        \subsubsection{DNN Hyperparameters, input/output properties}
            Similar to previous work \cite{manoj2021adversarial}, we tested on two \gls{dnn} architectures: (a) \texttt{fc.small} contain fully connected layers with $[64, 32, 32, 32, 5]$ neurons and (b) \texttt{fc.medium} contain fully connected layers with $[512, 256, 128, 128, 5]$ neurons.
            Two networks have 6,981 and 202,373 trainable parameters, respectively, and employ ReLU as activation units.
            Both models are trained on $329,000$ data pairs for $50$ epochs, with a batch size of $1024$ and a learning rate of $3e^{-4}$ using the Adam optimizer.

            The input properties for verification are constructed based on the perturbation constraint from previous adversarial attackers \cite{manoj2021adversarial,santos2021universal}.
            We assume the input is in the $L_\infty$-norm, \eg, the adversarial perturbation constraint is given as $[\eta_L, \eta_U] := [-\eta_0, \eta_0]^{J \times K \times 2}$.
            The level of perturbation $\eta_0$ is a scalar with values from $\{0.01, 0.1, 1, 10, 100\}$ (meter(s)) which correspond to distance perturbation of $1$ cm to $100$ m.
            The output properties, which are the maximum optimality gap $\zeta$, are a scalar and take values from $\{0.001, 0.005, 0.01, 0.05, 0.1\}$, which guarantee to retain 90\% to 99.9\% of the optimally trained model.
            For each property, we test the local robustness of 30 randomly drawn set of \gls{ue}s input locations from the test dataset.
            Also, we need to test $J=4$ models for $J=4$ cells.
            Thus, the total number of verification properties for the general scenario is $3,000$ properties.


        \subsubsection{Evaluation Metrics}

            \textit{$L_1$-distance between lower and upper output bound:}
            We measure the $L_1$-based distance between the lower bound $\rho_L$ and upper bound $\rho_U$ computed by abstraction-based \gls{nnv} \cite{singh2019abstract,xu2020automatic}.
            A smaller $L_1$-distance indicates tighter bounds and more predictable model behavior.

            \textit{Unsatisfiable percentage:} We report the percentage of \texttt{UNSAT} properties, averaged across multiple test inputs $x^{(n)}$.
            An \texttt{UNSAT} result indicates that the verification tool successfully proves that the \gls{dnn} model maintains the required optimality gap $\zeta$ despite input perturbations within the specified bounds $[\eta_L, \eta_U]$.

    \subsection{Robustness Analysis}
        We trained two architectures, \texttt{fc.medium} and \texttt{fc.small}, on a publicly available dataset to reproduce previous work~\cite {sanguinetti2018deep, manoj2021adversarial, manoj2022downlink}, which map \gls{ue} locations to maximum product \gls{sinr} power allocation.
        Under normal operational conditions without adversarial perturbations, both models achieve competitive performance with average throughput per \gls{ue} of approximately 30 Mbit/s or about 1.5bit/s/Hz, which corresponds to data from Fig. 2a in~\cite{sanguinetti2018deep}.

        \begin{table}[]
            \vspace{0.06in}
            \centering
            \begin{tabular}{@{}llllll@{}}
                \toprule
                \multirow{2}{*}{} & \multicolumn{5}{c}{Perturbation level $\eta_0$} \\ \cmidrule(l){2-6}
                 & \multicolumn{1}{c}{$\pm$ 1cm} & \multicolumn{1}{c}{$\pm$ 10cm} & \multicolumn{1}{c}{$\pm$ 1m} & \multicolumn{1}{c}{$\pm$ 10m} & $\pm$ 100m \\ \midrule
                \multicolumn{1}{c}{\texttt{fc.small}}  & 0.002  & 0.015  & 0.153  & 1.560  & 22.610  \\
                \multicolumn{1}{c}{\texttt{fc.medium}}  & 0.007  & 0.068  & 0.688  & 7.664  & 341.999  \\ \bottomrule
            \end{tabular}
            \caption{The average $L_1$-distance between $\rho_L$ and $\rho_U$ (mW). }
            \label{tab:l1}
        \end{table}
        \autoref{tab:l1} presents the output bound analysis for both \gls{dnn} architectures across different perturbation levels. The results demonstrate the relationship between input uncertainty and the bounds on output power allocation.
        As the perturbation level $\eta_0$ increases, the L1-distance between lower and upper bounds grows exponentially for both models. This behavior aligns with our expectations, where larger input perturbations lead to wider output ranges.
        However, despite having identical network depth, the \texttt{fc.medium} model consistently produces looser bounds compared to \texttt{fc.small}.
        This occurs because the wider architecture, with more neurons per layer, introduces additional complexity and nonlinearity, making abstraction-based verification more conservative in its bound estimates.

        \begin{table*}[t]
            \vspace{0.06in}
            \centering
            \begin{minipage}[t]{0.48\textwidth}
                \centering
                \begin{tabular}{@{}clllll@{}}
                    \toprule
                    \multicolumn{1}{l}{\multirow{2}{*}{}} & \multicolumn{5}{c}{Perturbation level $\eta_0$} \\ \cmidrule(l){2-6}
                    \multicolumn{1}{l}{} & $\pm$ 1cm & $\pm$ 10cm & $\pm$ 1m & $\pm$ 10m & $\pm$ 100m \\ \midrule
                    $\zeta=0.001$ & 100.0  & 100.0  & 6.7  & 0.0  & 0.0  \\
                    $\zeta=0.005$ & 100.0  & 100.0  & 80.0  & 0.0  & 0.0  \\
                    $\zeta=0.01$ & 100.0  & 100.0  & 100.0  & 0.0  & 0.0  \\
                    $\zeta=0.05$ & 100.0  & 100.0  & 100.0  & 70.0  & 0.0  \\
                    $\zeta=0.1$ & 100.0  & 100.0  & 100.0  & 100.0  & 0.0  \\ \bottomrule
                \end{tabular}
                \caption{The average percentage of \texttt{UNSAT} properties for model \texttt{fc.small}. }
                \label{tab:unsat:small}
            \end{minipage}
            \hfill
            \begin{minipage}[t]{0.48\textwidth}
                \centering
                \begin{tabular}{@{}clllll@{}}
                    \toprule
                    \multicolumn{1}{l}{\multirow{2}{*}{}} & \multicolumn{5}{c}{Perturbation level $\eta_0$} \\ \cmidrule(l){2-6}
                    \multicolumn{1}{l}{} & $\pm$ 1cm & $\pm$ 10cm & $\pm$ 1m & $\pm$ 10m & $\pm$ 100m \\ \midrule
                    $\zeta=0.001$ & 100.0  & 13.3  & 0.0  & 0.0  & 0.0  \\
                    $\zeta=0.005$ & 100.0  & 100.0  & 0.0  & 0.0  & 0.0  \\
                    $\zeta=0.01$ & 100.0  & 100.0  & 10.0  & 0.0  & 0.0  \\
                    $\zeta=0.05$ & 100.0  & 100.0  & 100.0  & 0.0  & 0.0  \\
                    $\zeta=0.1$ & 100.0  & 100.0  & 100.0  & 3.3  & 0.0  \\ \bottomrule
                \end{tabular}
                \caption{The average percentage of \texttt{UNSAT} properties for model \texttt{fc.medium}. }
                \label{tab:unsat:medium}
            \end{minipage}
        \end{table*}
        \autoref{tab:unsat:small} and \autoref{tab:unsat:medium} present the verification success rates for both model architectures across different combinations of perturbation levels $\eta_0$ and optimality gap requirements $\zeta$.
        The general trend reveals that 100\% \texttt{UNSAT} rates are consistently achieved when both the perturbation level $\eta_0$ is sufficiently small and the tolerable optimality gap $\zeta$ is sufficiently large.
        The \texttt{fc.small} model consistently exhibits higher verification success rates ($P_{\text{UNSAT}}$) compared to \texttt{fc.medium} across most experimental configurations.
        Particularly for the \texttt{fc.small} model, our verification results guarantee that input perturbations below $1$m can maintain 99\% of optimal performance ($\zeta \leq 0.01$) with 100\% confidence.

        These guarantees have significant implications for practical network deployment.
        \gls{bs} can implement anomaly-detection mechanisms to identify potential adversarial attacks by monitoring location reports for movement speeds exceeding specifications derived from our verification bounds.
        For instance, network operators can leverage the evaluation results when designing safe \gls{dnn}-based power allocation for massive \gls{mimo} networks, ensuring an optimality gap of no more than 10\% by constraining \gls{ue} movement to less than 10m between location update intervals.

        \begin{figure}[h]
            \centering
            \includegraphics[width=0.7\linewidth]{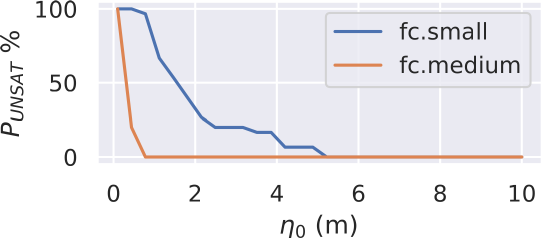}
            \caption{Compare two models \texttt{fc.small} and \texttt{fc.medium} in terms of the average percentage of \texttt{UNSAT} verification properties per each perturbation level $\eta_0$ and $\zeta=0.5\%$. }
            \label{fig:line1}
        \end{figure}
        \autoref{fig:line1} provides a more fine-grained analysis of the verification results presented in the previous tables, examining verification success rates across a continuous range of perturbation levels, $\eta_0 \in [0.1, 10]$ meters, with a fixed optimality gap tolerance of $\zeta=0.5\%$.
        The results demonstrate that \texttt{fc.small} maintains a high \texttt{UNSAT} rate for all perturbation levels up to $1$m, indicating strong provable robustness within this range.
        In contrast, \texttt{fc.medium} exhibits a sharp decline in \texttt{UNSAT} percentage, degrading rapidly as perturbation levels increase beyond 0.5m.
        Critically, both models lose all provable robustness guarantees when perturbation levels exceed 5m, with percentage of provable properties drop to 0\%.
        This establishes a clear operational boundary for safe deployment of these \gls{dnn}-based power control systems.

\section{Conclusion}
\label{sec:conclusion}




This work proposed a formal verification framework for \gls{dnn}-based power allocation in massive \gls{mimo}, providing optimality gap guarantee for \gls{dnn}-based power control against adversarial location perturbations via abstraction-based bounds and feasibility checking.
Evaluation results show that the smaller network architecture (\texttt{fc.small}) attains 100\% verification success up to 1m while preserving 99\% performance.
Future works will be extended to verifying different quality of service and scale verification to larger deep-learning models and more complex network settings.

\section*{Acknowledgement}

The authors thank the NICT AI R\&D Promotion Unit for providing cloud GPU credits.

\printbibliography
\end{NoHyper}

\end{document}